# HVPE-growth of Si-doped β-$Ga_2O_3$ on Sapphire: Influence of substrate offcut on structural and electrical properties

Sourav Sarker, Saleh Ahmed Khan, Ahmed Ibreljic, Anhar Bhuiyan [a)]
Department of Electrical and Computer Engineering, University of Massachusetts Lowell, Lowell, MA 01854, USA
[a)] Corresponding author Email: anhar_bhuiyan@uml.edu

**Abstract**

In this work, Si-doped β-$Ga_2O_3$ films were heteroepitaxially grown on sapphire substrates using halide vapor phase epitaxy (HVPE). The influence of sapphire offcut on growth kinetics, surface morphology, crystalline quality, and electrical transport properties was systematically investigated. Growth kinetics studies revealed a strong dependence of deposition rate on HCl flow, growth pressure, and Ga-source-to-substrate distance, with growth rates reaching up to 30 µm/hr. Increasing sapphire offcut angle from 0° to 8° promoted a transition from multidirectional growth to highly aligned terrace-dominated surfaces, reducing the surface roughness from 14.69 to 2.74 nm. The improved surface morphology was accompanied by enhanced crystalline quality, with phase-pure ($\bar{2}01$)-oriented β-$Ga_2O_3$ growth and a reduction in the rocking-curve full-width at half-maximum from 994 arcsec (0°) to 414 arcsec (6°). Electrical characterization of films grown on 6° offcut sapphire substrates at growth rates ranging from 0.67 to 7.30 µm/h yielded carrier concentrations ranging from $1.0 \times 10^{17}$ to $3.4 \times 10^{18}$ $cm^{-3}$. A maximum room-temperature hall mobility of 100 $cm^2/V{\cdot}s$ at a carrier concentration of $1.0 \times 10^{17}$ $cm^{-3}$ and a peak low-temperature mobility of 405 $cm^2/V{\cdot}s$ at 88 K were achieved for a 5.9-µm-thick film grown at 1.97 µm/h, representing the highest reported room- and low-temperature mobilities for HVPE-grown β-$Ga_2O_3$ on a sapphire substrate. Analysis of the temperature-dependent transport characteristics yielded donor activation energies of 35 and 90 meV together with a low acceptor concentration of $3 \times 10^{15}$ $cm^{-3}$, consistent with the improved crystalline quality achieved on the offcut sapphire substrates. These results demonstrate that HVPE is capable of producing high-quality β-$Ga_2O_3$ heteroepitaxial layers with good crystalline quality and carrier transport characteristics, providing a promising pathway for scalable β-$Ga_2O_3$ epitaxy on low-cost foreign substrates.

β-$Ga_2O_3$ has emerged as one of the most promising ultrawide bandgap semiconductors for next-generation high-power electronics due to its large bandgap (4.8-4.9 eV), high theoretical critical electric field approaching 8 MV/cm, and the availability of large-area melt-grown native substrates with diameters up to 6 inches[1, 2]. These attributes have enabled the demonstration of multi-kilovolt-class Schottky diodes[3-8], p-n heterojunction diodes[9-12], and field-effect transistors[13-18] with performance metrics that continue to advance toward the theoretical limits of the material system. Realization of high-performance β-$Ga_2O_3$ devices requires the development of thick, low-defect, and precisely controlled epitaxial layers. A variety of epitaxial growth techniques have been employed for β-$Ga_2O_3$, including metal-organic chemical vapor deposition (MOCVD) using cold-wall[19-38] and hot-wall[39-41] reactor configurations, halide vapor phase epitaxy (HVPE)[42-46], molecular beam epitaxy (MBE)[47-51], inorganic low-pressure CVD based on direct Ga-$O_2$ reactions[52-64], plasma-assisted Ga-$O_2$ deposition[65], and vacuum-free mist-CVD using either metalorganic or inorganic precursors[66]. HVPE has long been employed for the high-rate epitaxy of III–V semiconductors, including GaAs and GaN[67-70] and was later adopted for $Ga_2O_3$ growth[42-45, 71]. Its carbon-free halide chemistry enables high deposition rates and the growth of thick, high-quality layers with controlled low doping, making it particularly attractive for vertical power-device applications. For β-$Ga_2O_3$, these capabilities have been demonstrated primarily through homoepitaxy on native substrates, with high-quality films grown at rates ranging from several to tens of micrometers per hour[42-45]. Homoepitaxy eliminates film-substrate lattice and thermal-expansion mismatches and suppresses rotational-domain formation, thereby enabling excellent crystalline quality [22, 29, 30, 32, 33, 39, 42-45, 47, 52, 56, 60, 62, 72]. However, the relatively high cost of native substrates continues to motivate heteroepitaxy on lower-cost foreign substrates for scalable device applications. Among these substrates, sapphire is attractive because of its low cost, commercial availability in large wafer diameters, and excellent thermal stability[73].

Despite these advantages, heteroepitaxial growth of β-$Ga_2O_3$ on sapphire remains challenging because of the substantial lattice and thermal expansion mismatches between the film and substrate[54, 56, 74]. These mismatches often lead to rotational domains, increased defect densities, and degraded electrical transport properties. Previous studies have shown that substrate offcut can strongly influence epitaxial growth by modifying surface step density and promoting step-flow growth[54]. In β-$Ga_2O_3$ heteroepitaxy, intentional sapphire offcut has been shown to suppress competing growth orientations, improve terrace alignment, and enhance crystalline quality. However, most prior investigations of β-$Ga_2O_3$ growth on sapphire have focused on LPCVD-grown films[54, 56, 74], while studies employing HVPE for the heteroepitaxy of β-$Ga_2O_3$ films remain limited. Furthermore, the influence of sapphire offcut on the growth behavior, surface morphology, crystalline quality, and electrical transport properties of HVPE-grown β-$Ga_2O_3$ has not been systematically established. An earlier HVPE study comparing β-$Ga_2O_3$ growth on nominally on-axis and 5° offcut c-plane sapphire reported increased surface roughness for the offcut films[75], suggesting that further optimization of HVPE growth conditions, precursor flows, and reactor parameters is needed. In this work, we comprehensively investigate the heteroepitaxial growth of Si-doped β-$Ga_2O_3$ on c-plane sapphire substrates with different offcut angles using HVPE. The influence of substrate offcut on surface morphology, crystalline quality, and epitaxial evolution is systematically examined. Based on the optimized offcut condition, the effects of growth pressure and HCl flow rate on growth kinetics, carrier concentration, and electron mobility are subsequently evaluated. The results demonstrate the capability of HVPE to produce high-quality β-$Ga_2O_3$ films on sapphire with good structural and electrical properties and provide important insight into the growth and transport characteristics of HVPE-grown β-$Ga_2O_3$ on foreign substrate.

The β-$Ga_2O_3$ films were grown on nominally on-axis (0°), 2°, 6°, and 8° offcut c-plane sapphire substrates. Prior to growth, the substrates were sequentially cleaned in acetone, isopropyl alcohol (IPA), and deionized (DI) water, followed by nitrogen drying. No additional pre-growth annealing or surface treatment was performed. The HVPE growth process utilized high-purity metallic gallium (99.99999%) as the gallium source. HCl gas was introduced at flow rates ranging from 20 to 500 sccm, where it reacted with metallic gallium to generate volatile gallium chloride species for transport to the growth surface. Oxygen was supplied at flow rate ranging between 20 and 40 sccm as the oxidizing precursor for β-$Ga_2O_3$ formation. Silicon doping was achieved using $SiCl_4$, with flow rates varied between 1.25 and 12.5 sccm. The growth was carried out at a substrate temperature of 1050 °C under growth pressures ranging from low-pressure conditions (~5 Torr) to atmospheric pressure (760 Torr). The surface morphology of the as-grown films was characterized using field-emission scanning electron microscopy (FESEM, JEOL JSM-7401F) and atomic force microscopy (AFM, Park XE-100). The crystalline structure, phase purity, orientation, and structural quality of the films were evaluated using high-resolution X-ray diffraction (XRD) performed on a Rigaku SmartLab diffractometer equipped with Cu Kα radiation (λ = 1.5418 Å). Electrical transport properties were measured using the van der Pauw Hall configuration with an Ecopia HMS-5300 Hall measurement system. Temperature-dependent Hall measurements were carried out over the temperature range of 80-350 K to investigate carrier transport behavior and the temperature dependence of the electrical transport properties.

Figure 1 shows the influence of HCl flow rate, growth pressure, and source-to-substrate distance on the growth kinetics of HVPE-grown β-$Ga_2O_3$. As shown in Figure 1(a), the growth rate increases with increasing HCl flow for all Ga-source-to-substrate distances investigated, indicating that the supply of gallium chloride growth species strongly influences the deposition process. The growth rate exhibits a pronounced dependence on source-to-substrate distance, with shorter distances producing higher deposition rates. For a source-to-substrate distance of 3 cm, the growth rate increased from 15 μm/hr at 200 sccm HCl to 30 μm/hr at 300 sccm HCl. In comparison, at 3.5 cm the growth rate increased from 2.66 μm/hr at 80 sccm to 17.95 μm/hr at 300 sccm. The influence of reactor pressure on growth kinetics was further investigated at a fixed source-to-substrate distance of 5.5 cm and an HCl flow rate of 80 sccm, as shown in Figure 1(b). Increasing the growth pressure from 84 Torr to atmospheric pressure resulted in a substantial increase in growth rate from 0.12 to 2.66 μm/hr.

The surface morphology of the β-$Ga_2O_3$ films was investigated using FESEM, as shown in Figure 2. The films were grown under identical conditions using 90 sccm HCl, 20 sccm $O_2$, and 220 sccm Ar at atmospheric pressure for 1 h, resulting in a film thickness of ~8 μm. A pronounced evolution in surface morphology is observed as the sapphire offcut angle increases from 0° to 8°. The film grown on the nominally on-axis sapphire substrate [Fig. 2(a)] exhibits irregular elongated surface features with multiple in-plane orientations. The presence of intersecting domains and discontinuous surface structures indicates the absence of a dominant growth direction and suggests that nucleation and lateral growth occur along multiple crystallographic orientations. Such morphology is commonly observed during heteroepitaxial growth on nominally flat substrates where the density of surface steps is insufficient to effectively guide adatom incorporation [54, 74]. For the 2° offcut substrate [Fig. 2(b)], the surface morphology becomes noticeably more ordered. The elongated surface features exhibit a preferred alignment, indicating that the substrate miscut begins to influence growth directionality. The increased density of atomic steps introduced by the substrate offcut provides energetically favorable incorporation sites for migrating species, promoting more uniform lateral growth and reducing the occurrence of randomly oriented domains. Nevertheless, substantial variations in terrace width and surface roughness remain visible, suggesting that multiple growth modes still contribute to surface evolution. A significant change in morphology is observed for the 6° offcut sample [Fig. 2(c)]. The surface becomes dominated by highly aligned parallel terrace structures extending over large lateral distances with minimal interruption. The reduction in misaligned domains and surface discontinuities indicates a transition toward a more ordered growth regime. The long-range terrace alignment suggests enhanced adatom transport and more uniform incorporation at step edges, resulting in improved lateral growth uniformity. Similar morphological characteristics are observed for the 8° offcut sample [Fig. 2(d)], where the surface remains anisotropic and dominated by continuous terrace structures. Compared with the lower-offcut samples, both the 6° and 8° films exhibit improved surface ordering, indicating that the increased step density associated with larger substrate miscuts strongly influences the growth kinetics and promotes directional growth.

To quantitatively assess the surface morphology, AFM measurements were performed, as shown in Figure 3. For reference, AFM images of the sapphire substrates with different offcut angles are provided in Fig. S1 of the Supplementary Material, showing RMS roughness values of 0.23-0.58 nm. The epilayer AFM results are consistent with the SEM observations and reveal a systematic reduction in surface roughness with increasing substrate offcut. The film grown on the nominally on-axis substrate [Fig. 3(a)] exhibits large height variations associated with intersecting terrace structures and irregular faceted features. The corresponding root-mean-square (RMS) roughness is 14.69 nm, indicating substantial surface nonuniformity. For the 2° offcut sample [Fig. 3(b)], the RMS roughness decreases to 7.93 nm, reflecting a significant improvement in surface smoothness. Although elongated terrace structures become more evident, localized height variations remain present across the surface. Further increasing the offcut angle to 6° results in a reduction in roughness. The AFM image shown in Figure 3(c) reveals highly aligned terrace structures, yielding an RMS roughness of only 2.87 nm. The terraces extend continuously across the scanned area and exhibit a high degree of uniformity. A similarly smooth surface is observed for the 8° offcut sample [Fig. 3(d)], which exhibits the lowest RMS roughness of 2.74 nm. To further quantify the surface-height variations, AFM line profiles were obtained across representative terrace structures for all substrate offcut angles, as shown in Fig. S2 of the supplementary material. The line profiles reveal substantially smaller height variations for the 6° and 8° offcut samples than for the 0° and 2° samples, with the variations remaining within a few tens of nanometers, consistent with their lower RMS roughness.

To further investigate the influence of growth pressure on the surface morphology, FESEM measurements were performed on films grown on 6° offcut sapphire substrates under different pressures, as shown in Figure 4. The surface morphology changes significantly as the growth pressure increases. The film grown at 6 Torr [Fig. 4(a)] exhibits a surface composed of narrow elongated features and irregular terrace structures. Although the offcut substrate promotes a preferred growth direction, the terrace arrangement remains less uniform and contains a high density of morphological discontinuities. Increasing the growth pressure to 84 Torr [Fig. 4(b)] results in a substantial improvement in surface ordering. Well-defined terraces become more apparent and extend over larger lateral distances, indicating enhanced surface diffusion and more uniform incorporation of growth species. The terrace edges become smoother and exhibit a greater degree of parallel alignment compared with the low-pressure sample. A further increase in pressure to 760 Torr [Fig. 4(c)] produces more uniform morphology. The surface is dominated by highly aligned and continuous terrace structures, suggesting that higher-pressure conditions promote enhanced lateral propagation of atomic steps.

The same 8-μm thick β-$Ga_2O_3$ offcut-series samples characterized by FESEM and AFM in Figures 2 and 3 were also analyzed by high-resolution XRD, as shown in Figure 5. The ω-2θ scan obtained from the β-$Ga_2O_3$ film grown on a 6° offcut substrate, shown in Figure 5(a), is dominated by the β-$Ga_2O_3$ ($\bar{2}01$), ($\bar{4}02$), and ($\bar{8}04$) reflections, confirming the formation of phase-pure monoclinic β-$Ga_2O_3$ with strong preferential orientation along the ($\bar{2}01$) direction. No diffraction peaks associated with secondary $Ga_2O_3$ phases are detected. The effect of substrate offcut on crystalline quality was evaluated through rocking-curve measurements of the β-$Ga_2O_3$ ($\bar{4}02$) reflection, as shown in Figure 5(b). A systematic narrowing of the diffraction peak is observed as the substrate offcut angle increases, indicating improved crystallographic alignment. The broad diffraction profile observed for the nominally on-axis sample suggests a larger distribution of crystallographic tilt and a higher density of structural imperfections. In contrast, the higher-offcut samples exhibit significantly narrower rocking curves, reflecting improved crystalline coherence throughout the epilayer. The extracted rocking-curve full width at half maximum (FWHM) values are summarized in Figure 5(c). The FWHM decreases from 994 arcsec for the on-axis sample to 573 arcsec for the 2° offcut sample and reaches a minimum value near 414 arcsec for the 6° offcut sample. This improvement in crystalline quality is consistent with the enhanced surface ordering and terrace alignment observed in the SEM and AFM analyses. The lowest FWHM obtained for the 6° offcut sample suggests that this substrate orientation provides the most favorable growth conditions for achieving high crystalline quality under the present HVPE growth conditions. To further evaluate the dependence of crystalline quality on growth rate and film thickness, rocking-curve measurements were also performed on β-$Ga_2O_3$ films grown on 6° offcut sapphire at rates of 1.97, 7.30, and 10 µm/h, as shown in Fig. S3 of the supplementary material. The corresponding film thicknesses were 5.9, 7.3, and 30 µm, respectively. The FWHM increased from 399 arcsec at 1.97 µm/h to 505 arcsec at 7.30 µm/h and 669 arcsec at 10 µm/h, suggesting a trade-off between rapid thick-film growth and crystalline quality under the investigated conditions.

To benchmark the crystalline quality achieved in this work, the measured ω-rocking-curve FWHM values were compared with literature reports of β-$Ga_2O_3$ grown on sapphire using different epitaxial techniques and with previous HVPE results. Representative values include 256 arcsec for ~0.125-µm-thick MBE film [76], 886 arcsec for a ~0.387-µm-thick MOCVD film [77], ~1200 to more than ~7200 arcsec for previously reported HVPE-grown films [71, 75], and ~4072 arcsec for a 0.338-µm-thick mist-CVD β-$Ga_2O_3$ layer grown with an intermediate buffer layer [78]. Our recent LPCVD study achieved a FWHM as low as 293 arcsec for an 11.47-µm-thick film grown on 6° offcut c-plane sapphire [56]. In the present HVPE study, the 8-µm-thick offcut-series films exhibited a reduction in FWHM from 994 arcsec for the on-axis sample to 414 arcsec for the 6° offcut sample, while a minimum FWHM of 399 arcsec was obtained for a 5.9-µm-thick film grown at 1.97 µm/h. This value is substantially narrower than most previously reported β-$Ga_2O_3$ films on sapphire and represents, to the best of our knowledge, the lowest rocking-curve FWHM reported for HVPE-grown β-$Ga_2O_3$ directly on sapphire.

The room-temperature electrical transport properties of Si-doped β-$Ga_2O_3$ films grown on 6° offcut sapphire substrates are listed in Table 1. By varying the HVPE growth conditions, carrier concentrations ranging from $1.0 \times 10^{17}$ to $3.4 \times 10^{18}$ $cm^{-3}$ and electron mobilities ranging from 21 to 100 $cm^2/V\cdot s$ were achieved. Among the investigated films, Sample #5, with a thickness of 5.9-μm, exhibited the highest room-temperature mobility of 100 $cm^2/V\cdot s$ at a carrier concentration of $1.0 \times 10^{17}$ $cm^{-3}$. For Samples #1 and #2, increasing the $SiCl_4$ flow from 1.25 to 10 sccm was accompanied by a modest increase in carrier concentration from $1.5 \times 10^{18}$ to $2.0 \times 10^{18}$ $cm^{-3}$. However, because the growth pressure and source-to-substrate distance also differed between these samples, the carrier-concentration response cannot be attributed solely to the $SiCl_4$ flow and may reflect the combined influence of precursor delivery and transport conditions. The concurrent decrease in mobility from 77 to 35 $cm^2/V\cdot s$ may indicate increased impurity- or defect-related scattering. A more direct comparison is provided by Samples #3 and #4, which were grown at the same pressure and with identical precursor flows. Decreasing the source-to-substrate distance from 5.7 to 4.0 cm increased the growth rate from 0.73 to 0.83 μm/h, while the carrier concentration increased from $5.0 \times 10^{17}$ to $3.4 \times 10^{18}$ $cm^{-3}$, suggesting that dopant incorporation is highly sensitive to precursor transport. At atmospheric pressure, Samples #5 and #6 were grown with identical Ar and $SiCl_4$ flows. The higher HCl and $O_2$ flows used for Sample #6 were accompanied by an increase in growth rate from 1.97 to 7.30 μm/h, despite relatively larger Ga-source-to-substrate distance. The carrier concentration increased from $1.0 \times 10^{17}$ to $2.4 \times 10^{17}$ $cm^{-3}$, whereas the mobility decreased from 100 to 21 $cm^2/V\cdot s$, suggesting a combined influence of precursor supply, growth kinetics, and carrier-scattering mechanisms. Residual Cl incorporation, which has been reported in HVPE-grown $\beta$-$Ga_2O_3$ particularly at higher growth rates [42, 44], may represent an additional consideration. However, evaluating its possible contribution remains beyond the scope of this study and will require further compositional analysis.

To investigate the carrier activation and scattering mechanisms, temperature-dependent Hall measurements were performed on Sample #5 over the temperature range of 80-350 K, as shown in Figure 6. The carrier concentration decreases from 1.0 × $10^{17}$ $cm^{-3}$ at room temperature to 1.8 × $10^{16}$ $cm^{-3}$ at 88 K, indicating progressive carrier freeze-out at lower temperatures. Simultaneously, the Hall mobility increases from 100 $cm^2/V{\cdot}s$ at room temperature to a peak value of 405 $cm^2/V{\cdot}s$ at 88 K, reflecting the suppression of thermally activated scattering processes. The high low-temperature peak mobility further supports the high electronic quality of the heteroepitaxial film. The experimental carrier concentration and mobility data were further analyzed using a charge-neutrality model combined with Boltzmann transport calculations [79-81], with details provided in the Supplementary material. The fitting reveals two donor populations with activation energies of 35 and 90 meV and concentrations of 1.39 × $10^{17}$ and 1.0 × $10^{16}$ $cm^{-3}$, respectively, as summarized in Table 2. The extracted acceptor concentration and threading dislocation density were relatively low, at 3 × $10^{15}$ $cm^{-3}$ and 2.2 × $10^{8}$ $cm^{-2}$, respectively. The calculated mobility components indicate that polar optical phonon scattering dominates at elevated temperatures, while impurity- and defect-related scattering mechanisms become increasingly important at lower temperatures [22, 23, 82]. The good agreement between the experimental data and the transport model indicates that the extracted donor concentrations, activation energies, and dislocation density effectively describe the carrier transport characteristics of the HVPE-grown β-$Ga_2O_3$ films.

To benchmark the electrical-transport performance achieved in this work, the Hall results were compared with representative reports of β-$Ga_2O_3$ heteroepitaxy on sapphire. Previous studies using MBE, MOCVD, and mist CVD have largely focused on structural and morphological properties, with reported film thicknesses of ~0.125-0.387 μm and provided limited Hall-transport data [76-78]. Our recent LPCVD study achieved a record-high room-temperature Hall of 149 $cm^2/V{\cdot}s$ at carrier concentration of $1.15 \times 10^{17}$ cm$^{-3}$ in a 10.15-μm-thick Si-doped β-$Ga_2O_3$ film on sapphire [56]. However, high-mobility Si-doped β-$Ga_2O_3$ grown by HVPE on sapphire has not previously been established [71, 75, 83]. In this present work, a 5.9-μm-thick HVPE film exhibited a room-temperature mobility of 100 cm$^2$/V·s at $1.0 \times 10^{17}$ cm$^{-3}$ and a peak mobility of 405 cm$^2$/V·s at 88 K. To the best of our knowledge, this is the first demonstration of high-mobility Si-doped β-$Ga_2O_3$ grown by HVPE directly on sapphire, and these are the highest room- and low-temperature Hall mobilities reported for such films.

In summary, HVPE grown high-quality Si-doped β-$Ga_2O_3$ heteroepitaxy on sapphire was demonstrated. Sapphire offcut was found to strongly influence surface evolution and crystalline quality, with the 6° offcut substrate providing the optimum growth surface. A room-temperature mobility of 100 cm$^2$/V·s and a peak mobility of 405 cm$^2$/V·s at 88 K were achieved, demonstrating excellent electron-transport properties in HVPE-grown β-$Ga_2O_3$ on sapphire. Temperature-dependent Hall analysis further revealed low acceptor compensation and a low threading dislocation density, consistent with good structural quality of the films. These findings provide critical insight into the growth, structure and transport characteristics of β-$Ga_2O_3$ heteroepitaxy and demonstrate the potential of HVPE as a scalable approach for producing high-quality β-$Ga_2O_3$ epitaxial layers on low-cost foreign substrates.

See the Supplementary Material for additional AFM surface maps and line profiles, AFM images of the sapphire substrates, growth-rate-dependent XRD rocking curves, and details of the charge-neutrality analysis, Boltzmann transport modeling, scattering-rate calculations, fitting procedures, and material parameters used to analyze the temperature-dependent Hall-effect measurements.

**Acknowledgement**

This work was supported by the National Science Foundation under Award ECCS-2501623 and Award ECCS-2532898 and Air Force Research Laboratory Regional Network-Mid-Atlantic (AFRL RNMA).

**Data Availability**

The data that support the findings of this study are available from the corresponding author upon reasonable request.

**Conflict of Interest**

The authors have no conflicts to disclose.

**Table 1**

Growth conditions and room-temperature electrical transport properties of HVPE-grown Si-doped β-$Ga_2O_3$ films on 6° offcut sapphire substrates.

| Sample | Pressure (Torr) | HCl Flow (sccm) | $Ar/O_2/SiCl_4$ Flow (sccm) | Source-to-Substrate Distance (cm) | Thickness (µm) | Growth Rate (µm $h^{-1}$) | Carrier Concentration ($cm^{-3}$) | Hall Mobility ($cm^2$ $V^{-1}$ $s^{-1}$) |
|---|---|---|---|---|---|---|---|---|
| #1 | 4.7 | 20 | 220/20/1.25 | 11.0 | 2.3 | 0.77 | $1.5 \times 10^{18}$ | 77 |
| #2 | 5.9 | 20 | 220/20/10 | 7.0 | 2.0 | 0.67 | $2.0 \times 10^{18}$ | 35 |
| #3 | 5.5 | 20 | 220/20/2 | 5.7 | 2.2 | 0.73 | $5.0 \times 10^{17}$ | 39 |
| #4 | 5.5 | 20 | 220/20/2 | 4.0 | 2.5 | 0.83 | $3.4 \times 10^{18}$ | 52 |
| #5 | 760 | 80 | 40/20/12.5 | 3.5 | 5.9 | 1.97 | $1.0 \times 10^{17}$ | 100 |
| #6 | 760 | 300 | 40/40/12.5 | 4.5 | 7.3 | 7.30 | $2.4 \times 10^{17}$ | 21 |

**Table 2**

Extracted donor parameters, acceptor concentration, and dislocation density obtained from fitting of the temperature-dependent Hall data.

| Sample No | $N_{D1}$ (cm⁻³) | $E_{D1}$ (meV) | $N_{D2}$ (cm⁻³) | $E_{D2}$ (meV) | $N_A$ (cm⁻³) | $N_{dis}$ (cm⁻²) |
|---|---|---|---|---|---|---|
| #5 | $1.39 \times 10^{17}$ | 35 | $1.0 \times 10^{16}$ | 90 | $3 \times 10^{15}$ | $2.2 \times 10^{8}$ |

**Figure 1**

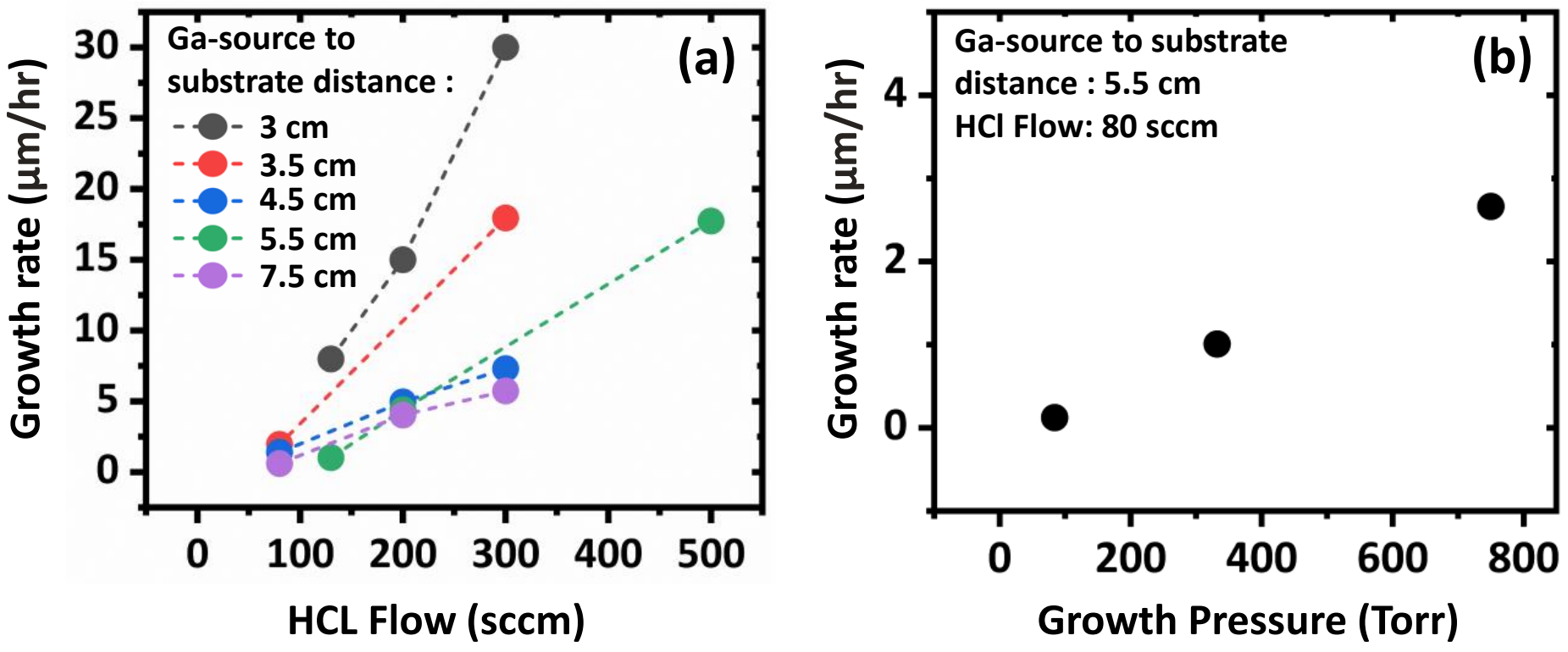

**Figure 1.** Dependence of β-$Ga_2O_3$ growth rate on (a) HCl flow rate for various Ga-source-to-substrate distances at atmospheric pressure and (b) growth pressure at a fixed Ga-source-to-substrate distance of 5.5 cm and HCl flow rate of 80 sccm.

**Figure 2**

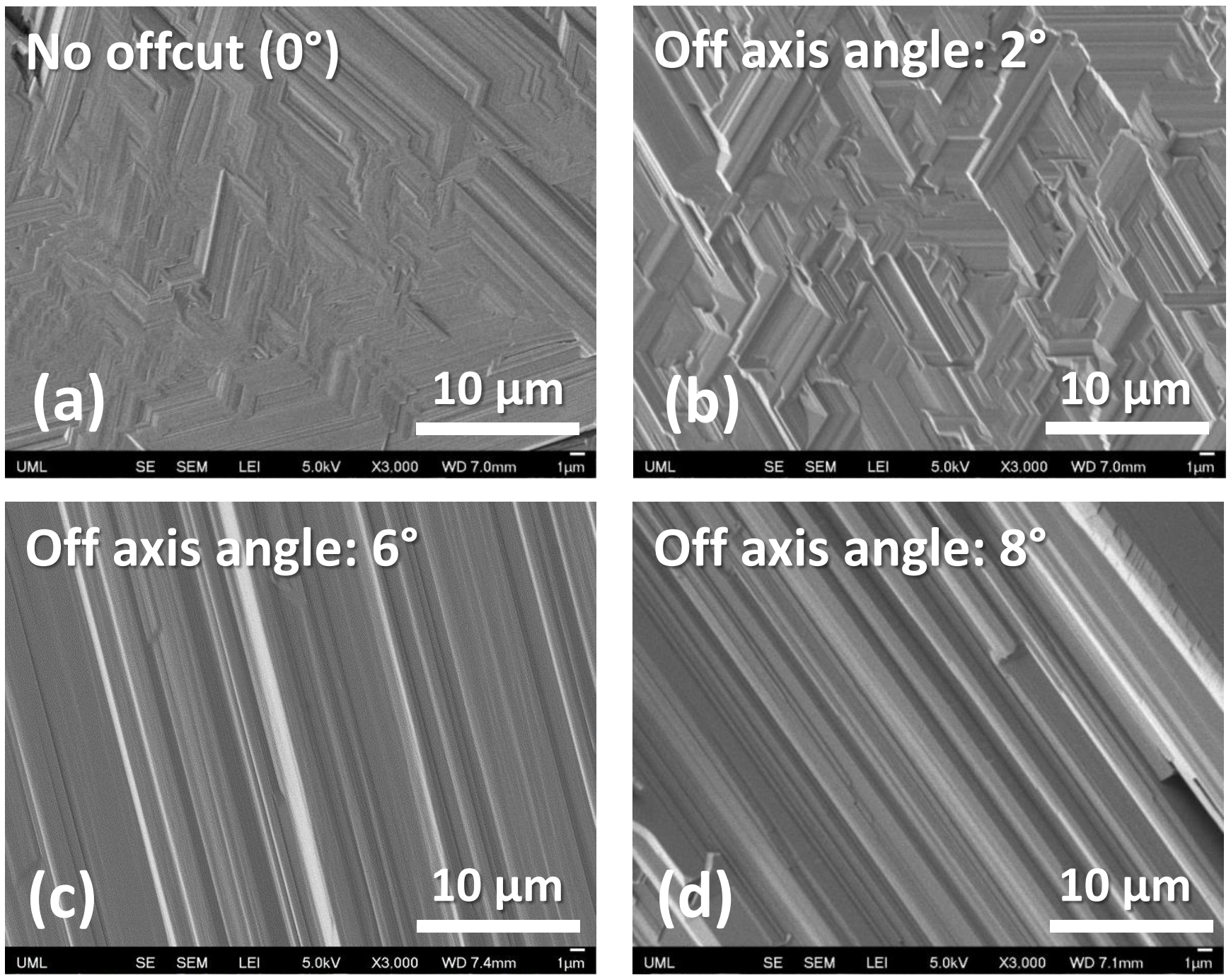

**Figure 2.** Surface view FESEM images of $\beta$-$Ga_2O_3$ films on c-plane sapphire substrates with offcut angles of (a) 0°, (b) 2°, (c) 6°, and (d) 8°.

**Figure 3**

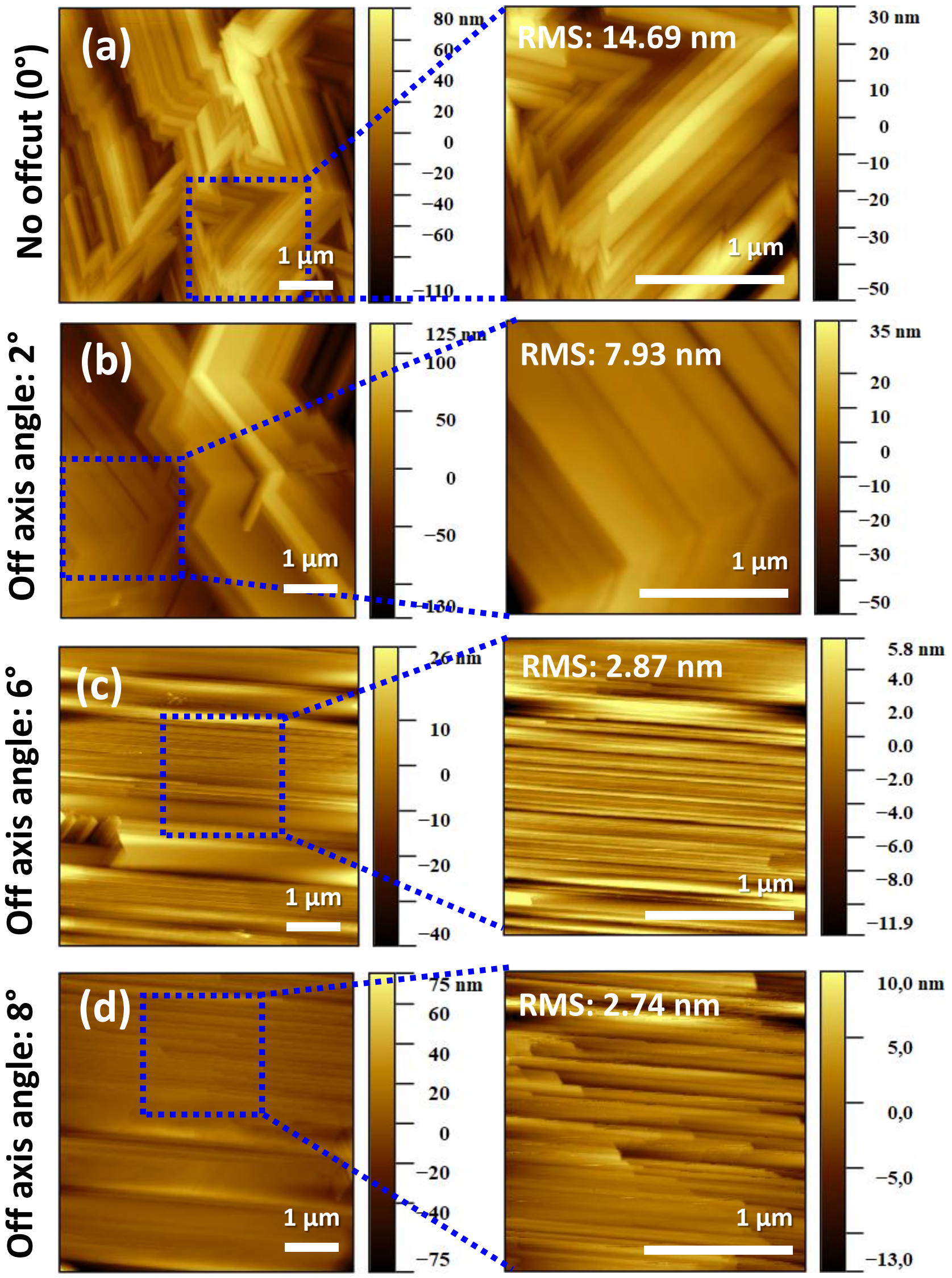

**Figure 3.** Surface AFM images of β-$Ga_2O_3$ films grown on sapphire substrates with offcut angles of (a) 0°, (b) 2°, (c) 6°, and (d) 8°.

**Figure 4**

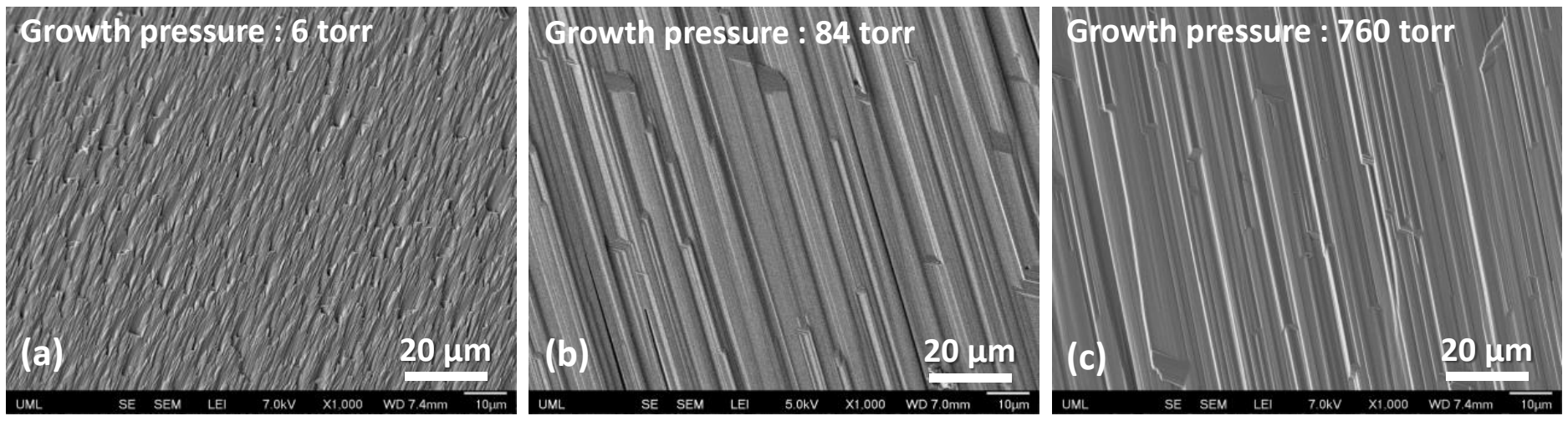

**Figure 4.** Surface-view FESEM images of β-$Ga_2O_3$ films on 6° offcut sapphire substrates grown at pressures of (a) 6 Torr, (b) 84 Torr, and (c) 760 Torr.

**Figure 5**

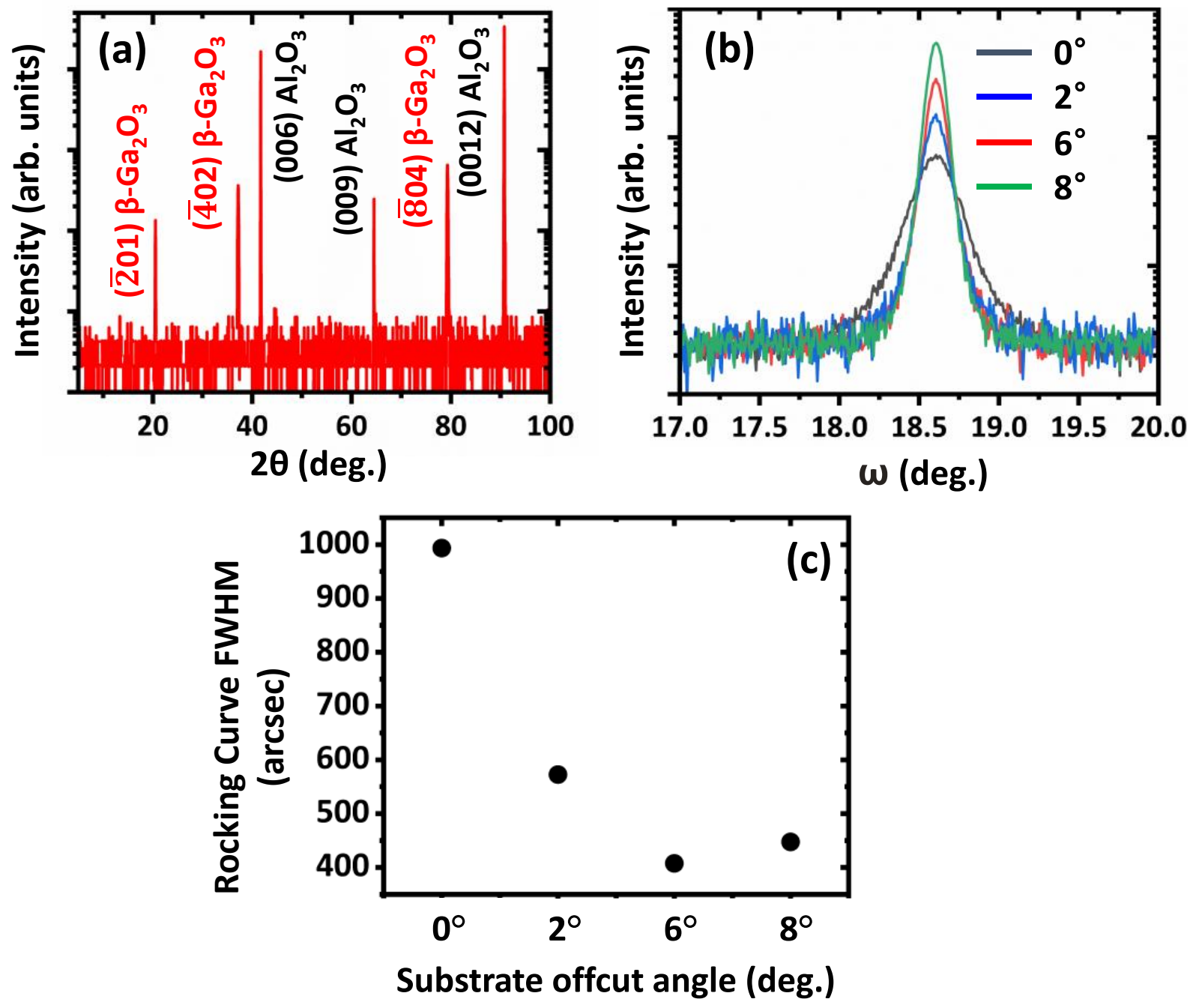

**Figure 5.** High-resolution XRD of β-$Ga_2O_3$ films on sapphire substrates. (a) ω-2θ scan of the β-$Ga_2O_3$ film grown on a 6° offcut sapphire. (b) Rocking curves of the β-$Ga_2O_3$ ($\bar{4}$02) reflection for different sapphire offcut angles. (c) Extracted rocking-curve FWHM as a function of substrate offcut angle.

**Figure 6**

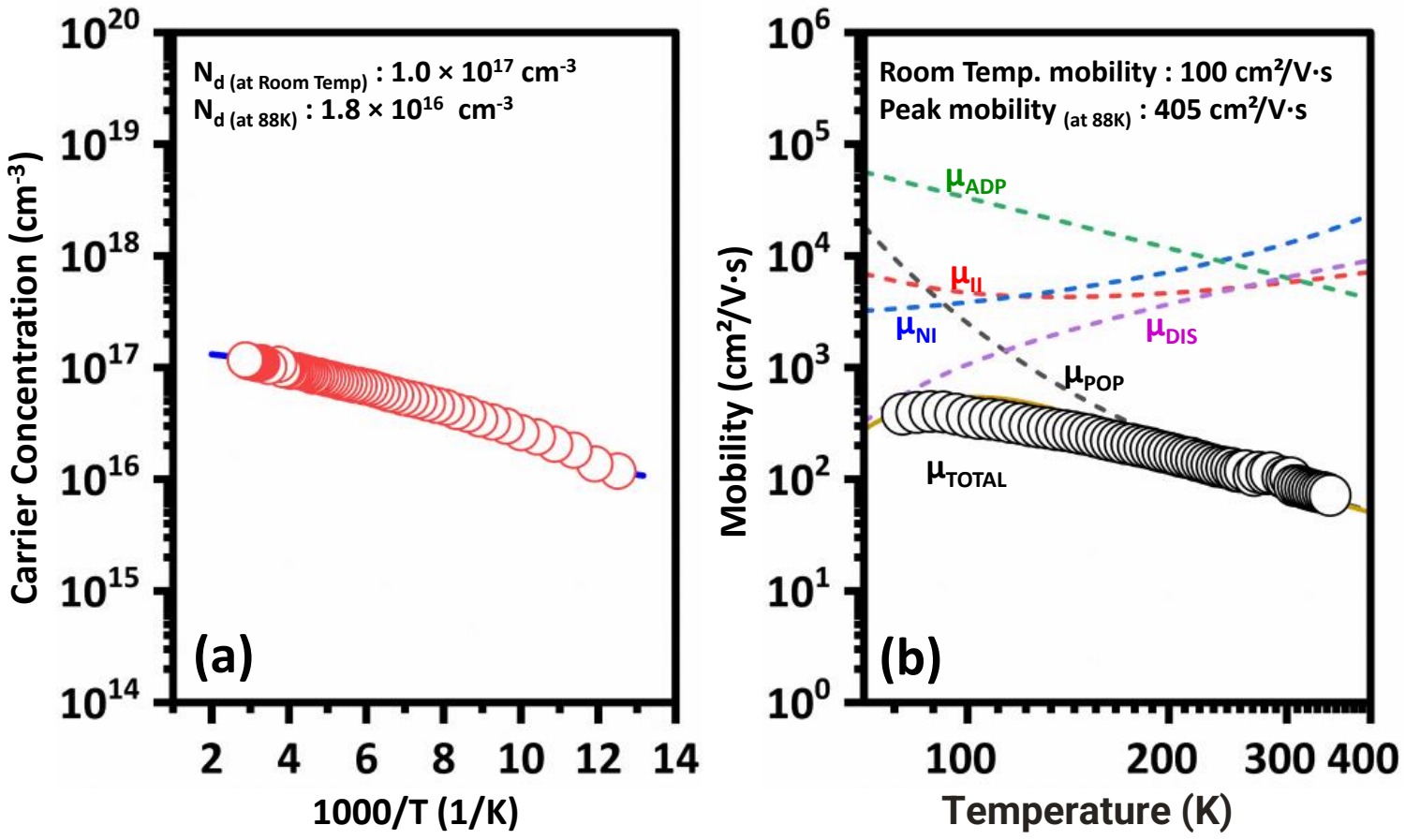

**Figure 6.** Temperature-dependent Hall transport characteristics of the Si-doped β-$Ga_2O_3$ film on a 6° offcut sapphire substrate. (a) Carrier concentration as a function of 1000/T together with the fitted charge-neutrality model. (b) Hall mobility as a function of temperature, showing the measured mobility and calculated contributions from polar optical phonon scattering, acoustic deformation potential scattering, ionized impurity scattering, neutral impurity scattering, and dislocation scattering.